\documentclass[final,3p,times]{elsarticle}
 \biboptions{comma,sort&compress}
\usepackage{here}
 \usepackage{graphicx}
  \usepackage{epsfig}

\usepackage{amsmath,bm}

\def\beq{\begin{equation}}
\def\eeq{\end{equation}}
\def\bea{\begin{eqnarray}}
\def\eea{\end{eqnarray}}

\def\lesssim{\ \hbox{\raise 2pt \hbox{$<$} \kern -13pt
                     \lower 3pt \hbox{$\sim$}}\ }
\def\greatersim{\ \hbox{\raise 2pt \hbox{$>$} \kern -13pt
                     \lower 3pt \hbox{$\sim$}}\ }
\def\frac#1#2{ {{#1} \over {#2} }}

\input epsf.tex
\def\desepsf(#1 width #2){\epsfxsize=#2 \epsfbox{#1}}

\newcommand{\as}{\alpha_\mathrm{s}}

\begin{document}

\begin{frontmatter}

\hspace*{13.0 cm} {\small 
DESY 16-174} \\
\vspace*{1.4 cm} 
\title{Soft-gluon resolution scale in QCD evolution equations}
\author[label1,label2,label3]{F. Hautmann}
 \address[label1]{Rutherford Appleton Laboratory, Chilton OX11 0QX}
  \address[label2]{Theoretical Physics  Department, University of Oxford,  Oxford OX1 3NP}
\address[label3]{Elementaire Deeltjes Fysica, Universiteit Antwerpen, B 2020 Antwerpen}
 \author[label4]{H. Jung}
  \address[label4]{Deutsches Elektronen Synchrotron,  D 22603 Hamburg }

 \author[label4]{A. Lelek}

\author[label5]{V. Radescu} 
\address[label5]{CERN,  CH-1211 Geneva 23 }

 \author[label4]{R. \v{Z}leb\v{c}\'{i}k}

\begin{abstract}
QCD evolution equations can 
be recast in terms of 
parton branching processes.  
We present a new 
numerical solution of the 
equations. 
We show that this parton-branching 
solution can be applied  
to analyze infrared contributions 
to evolution, order-by-order in the strong coupling $\alpha_s$, as a 
function of the soft-gluon resolution scale parameter.  We examine the 
cases of transverse-momentum 
ordering and angular ordering. 
We illustrate that this approach can 
be used to treat 
distributions  which depend  
both  on 
longitudinal  and  on 
transverse momenta. 
\end{abstract}

\end{frontmatter}

The evolution of QCD parton cascades is an essential element of 
theoretical predictions for 
production  
  processes  with high  momentum transfer 
 at  high-energy colliders.  
It 
has  been realized since long 
that realistic 
predictions for collider processes   
require taking  into account   contributions to  QCD  evolution 
not only from collinear parton radiation, associated with the renormalization group behavior at high momenta, but also from soft 
gluon radiation, including   its color coherence 
properties~\cite{Webber:1986mc,bcm83,Dokshitzer:1987nm,Marchesini:1987cf,Catani:1990rr}. 
Infrared radiation contributions   
are controlled by a finite 
resolution scale $\delta$ of order 
$\delta \sim {\cal O} (  \Lambda_{\rm{QCD}} /  
\mu )$, where $\mu$ is the hard  scattering scale and $  \Lambda_{\rm{QCD}} \approx 
1$~fm$^{-1}$ is the natural scale 
of strong interactions. 

In this paper we study the effects 
of soft-gluon emission and the 
soft-gluon resolution scale in 
cases in which not only 
longitudinal-momentum degrees of freedom but also 
transverse-momentum degrees 
of freedom 
are necessary for reliable theoretical  predictions. We address  the issue of  taking  into account simultaneously soft gluon radiation, with light-cone momentum fraction $z \to 1$,  and 
transverse momentum 
${\bf q}_\perp$ recoils in the parton branchings along the QCD cascade. 
This is relevant  for instance in  multiple-scale  problems,  
 such as the high 
invariant-mass  region of heavy  particle spectra and the high-energy  limit of hadroproduction processes,  
where   transverse-momentum  dependent (TMD) factorization  theorems  apply (see  e.g.~\cite{Angeles-Martinez:2015sea}  for a recent review).  

While analytic 
resummation methods exist, based on these theorems,  for sufficiently inclusive variables  such as, e.g., 
heavy-boson transverse   spectra,    
the viewpoint in this work is to 
aim at a formulation by which one 
could also treat exclusive components 
of the final states. 
Parton-shower algorithms do provide 
such a formulation, and are widely 
used as an effective  alternative 
(though limited in accuracy) to 
analytic resummation methods. 
While great progress 
has been achieved in the last 
decade on matching and merging 
methods~\cite{Hoeche:2011zz}   
to combine parton showers with 
perturbative calculations 
through next-to-leading order,  
several  open  questions still 
remain, 
both conceptual and technical,   
on the appropriate use of parton   
distribution functions in parton   
showers~\cite{Nagy:2014oqa}     
and on the treatment of the   
shower's transverse momentum   kinematics~\cite{Dooling:2012uw}.   

In the context of analytic  methods, 
the behavior of parton distributions near the endpoint $ z \to 1$ 
motivates the use of infrared subtractive techniques  which lead to 
a  generalization of the ``plus"  
distribution~\cite{fh07}   
including   transverse degrees of 
freedom. 
In this paper we describe a new 
calculation,  based on the unitarity 
method~\cite{Webber:1986mc} to  
recast evolution equations in terms 
of Sudakov form factors and real 
emission kernels, and present a 
study of the soft-gluon resolution 
scale in the cases  of inclusive 
parton distributions and of 
transverse-momentum dependent 
parton distributions. 
We analyze 
 different ordering variables, 
including transverse momentum ordering and angular ordering. 
The method set up  
in this paper can be applied 
systematically order-by-order in the strong coupling $\alpha_s$, at  
 leading order as well as at next-to-leading and higher  orders. 
In this article we present the basic 
results and numerical 
leading-order applications. Details of the method and applications including 
next-to-leading order will be 
presented elsewhere~\cite{prepa}.   

We start from
the  renormalization group  evolution of parton distribution 
functions~\cite{dglapref1,dglapref2,dglapref3} 
\begin{equation}
\label{evapp}
\frac{\partial \;{\widetilde f}_a(x,\mu^2)}{\partial \ln \mu^2} = \sum_b \int_x^1 {dz} \;
P_{ab}(\as(\mu^2),z) \;{\widetilde f}_b({x/z},\mu^2) \;\;.  
\end{equation}
where $
{\widetilde f}_{a} (x,\mu^{2}) \equiv x 
f_{a}(x, \mu^{2})$ 
 are 
momentum-weighted 
 parton distributions   for $a = 1 , \dots , 2 N_f + 1 $ species of partons   
(with $N_f$  the number of quark  flavors)   
 as functions of longitudinal 
momentum fraction $x$ and  
evolution mass scale $\mu$, and  
$P_{ab} (\as ,z)$ are 
splitting functions,  computable as  perturbation 
series expansions  in powers of the 
strong coupling $\alpha_s$.

We  classify the singular 
behavior  
 of the splitting functions 
$P_{ab} (\as ,z)$ 
for $z \to 1$ 
 according to the decomposition 
\begin{equation}
P_{ab} (\as ,z) = 
D_{ab} (\as) \delta ( 1 - z ) 
+ K_{ab} (\as) \ 
{ 1 \over { ( 1 - z )_+ }} 
+ R_{ab} (\as ,z)
\;, 
\label{decompPab}
\end{equation}
where the plus-distribution 
${ 1 / { ( 1 - z)_+ }} $ 
is defined for any test function 
$\varphi$ as 
\begin{equation}
\int_0^1 { 1 \over { ( 1 - z)_+ }}  \ 
\varphi ( z )  \ dz  = 
\int_0^1 { 1 \over {  1 - z }}  \ [ 
\varphi ( z )  - \varphi ( 1 ) ] \  dz
\;. 
\label{plusdistrdef}
\end{equation}
Eq.~(\ref{decompPab}) decomposes 
the splitting functions into the 
 $\delta (1-z)$ distribution, 
the ${ 1 / { ( 1 - z)_+ }} $ distribution, 
and the function  $R (\as ,z)$ which 
contains logarithmic terms in 
 $\ln (1-z)$ and analytic terms for 
$z \to 1$.  The 
$\delta (1-z)$ 
and ${ 1 / { ( 1 - z)_+ }} $
contributions to  
splitting functions    are 
diagonal in flavor, 
\begin{equation}
D_{ab} (\as) = \delta_{a b } 
d_a (\as)  
\, , \;\; 
K_{ab} (\as) = 
\delta_{a b } 
k_a (\as)  
\,  
\label{flavdiag}
\end{equation}
(no summation  
over repeated indices).  
The constants 
$d_{a}$ 
and $k_{a}$  and the functions 
$ R_{ab} $ 
in Eq.~(\ref{decompPab}) 
can be expanded 
in powers of $\alpha_s$. 
The two-loop expansions for the 
constants $d_{a}$ 
and $k_{a}$ may be obtained 
from~\cite{cfpref1,cfpref2} and  
read 
\begin{eqnarray}
 d_q &=& 
 \frac{3 \as C_F}{4\pi} 
+  \left( \frac{\as}{2\pi}
\right)^{2} \left[ 
C_F^2 \left( {3 \over 8} -  
{\pi^2 \over 2} + 6 \ \zeta (3) \right) 
+ 
C_F C_A \left( 
{17 \over 24} +   
{{11 \pi^2} \over 18} - 3 \ \zeta (3)
 \right) 
-  C_F T_R N_f \left( 
{1 \over 6} +   
{{2 \pi^2} \over 9} 
 \right)  \right] + {\cal O } 
\left( \as^3
\right)
 \; , 
\nonumber\\ 
 d_g &=&  \frac{\as}{2\pi} 
\left(  {11 \over 6} \, C_A - 
{2 \over 3} \, T_R \, N_f
\right)  
+  \left( \frac{\as}{2\pi}
\right)^{2} \left[
C_A^2 \left( {8 \over 3} + 3 \ 
 \zeta (3) \right) 
- {4 \over 3} \ C_A T_R N_f 
-  C_F T_R N_f   \right] + {\cal O } 
\left( \as^3 
\right) \; , 
\nonumber\\ 
k_q &=&   \frac{ \as C_F}{\pi} 
+  \frac{\as^2 C_F}{2\pi^2}
\ \Gamma + {\cal O } 
\left( \as^3 
\right)
   \; , 
\;\; k_g = \frac{ \as C_A}{\pi} 
+  \frac{\as^2 C_A}{2\pi^2}
\ \Gamma + {\cal O } 
\left( \as^3 
\right)
   \; , 
\;\;   \Gamma  \equiv   C_A \left( 
{ 67 \over 18 } - {\pi^2 \over 6} 
\right) - T_R N_f {10 \over 9 } \; ,
\label{twoloopconst}
\end{eqnarray}
where $
C_A = N_c$,  $C_F = (N_c^2 - 1) / ( 2 \, N_c)$, $T_R  = {1 \over 2}$ are 
 ${\rm{SU}}(N_c)$ color factors  ($N_c =3 $), 
and $\zeta$ is the Riemann zeta function. 
Analogously, explicit 
expressions for the 
functions $R_{ab}$ may 
be obtained 
from~\cite{cfpref1,cfpref2} 
through two loops.

In the physical picture of   Eqs.~(\ref{evapp}),(\ref{decompPab})   
a finite resolution scale 
in the transverse distance between emitted partons implies, by energy-momentum 
conservation,  that 
partons radiated with 
longitudinal momentum fractions closer to $ z = 1$ than a 
certain cut-off value, 
$ z > z_M $ with $ 1-z_M \sim 
{\cal O} ( \Lambda_{\rm{QCD}} / 
\mu )$,    
  cannot be resolved. 
Removing such radiative contributions from the evolution, on 
the other hand,  leads to a violation  of unitarity. 
The key idea of the parton branching 
method 
is to restore unitarity by  recasting 
 the evolution equations in terms of 
no-branching probabilities (Sudakov 
form factors) and 
real-emission branching 
 probabilities~\cite{Webber:1986mc,Sjo1986}.     

To this end, in this work we proceed 
in two steps, as follows. First, we introduce the resolution scale parameter $z_M $ into the evolution 
equations (\ref{evapp}) by splitting the 
integration range on the right hand side into the resolvable ($z < z_M$)  and non-resolvable ($z > z_M$) regions. We include terms through  
$ {\cal O } ( 1 - z_M)^0$ but neglect 
power-suppressed contributions 
$ {\cal O } ( 1 - z_M)^n$, $ n \geq 1$. 
(The details of this analysis will be 
given elsewhere~\cite{prepa}.) 
Further we use the momentum sum 
rule  
\begin{equation}
\label{momsum}
 \sum_c \int_0^1  z  
 \   P_{ca}(\as,z) \ dz = 0 
\;\; ({\rm{for}} \;\; {\rm{any}} \;\; a)  
\;  
\end{equation} 
to systematically 
eliminate $D$-terms in 
Eq.~(\ref{decompPab})
in favor of $K$- and $R$-terms. 
Then the evolution 
equations (\ref{evapp}) 
can be recast in integral form as 
\begin{eqnarray}
\label{sudint}
  {\widetilde f}_a(x,\mu^2) 
& = &  
S_a ( z_M, \mu^2 , \mu^2_0 ) \ 
 {\widetilde f}_a(x,\mu^2_0)  
+ 
\sum_b 
\int^{\mu^2}_{\mu^2_0} 
{{d \mu^{\prime 2} } 
\over \mu^{\prime 2} } 
{
{S_a ( z_M, \mu^2 , \mu^2_0 )} 
 \over 
{S_a ( z_M, \mu^{\prime 2} , 
\mu^2_0 ) }
}
\nonumber\\ 
& \times & 
\int_x^{z_M} {dz} \;
\left( 
K_{ab} (\as(\mu^{\prime 2})
) \ { 1 \over { 1 - z}} + 
R_{ab} (\as(\mu^{\prime 2})
,z) 
\right) 
\;{\widetilde f}_b({x/z},
\mu^{\prime 2})  
+ {\cal O } (1 - z_M) 
  \;    , 
\end{eqnarray}
where $S_a$ is the Sudakov form 
factor 
\begin{equation}
\label{suddef}
 S_a ( z_M, \mu^2 , \mu^2_0 ) = 
\exp \left[  -  \sum_b  
\int^{\mu^2}_{\mu^2_0} 
{{d \mu^{\prime 2} } 
\over \mu^{\prime 2} } 
 \int_0^{z_M} dz \  z 
\left( 
K_{ab} (\as(\mu^{\prime 2})
) \ { 1 \over { 1 - z}} + 
R_{ab} (\as(\mu^{\prime 2})
,z) 
\right)
\right] 
  \;  .  
\end{equation}  
Eqs.~(\ref{sudint}),(\ref{suddef}) 
take into account 
the effects of the $d$-terms in  
Eq.~(\ref{twoloopconst}) and 
 subtraction term in 
Eq.~(\ref{plusdistrdef}) implicitly, 
while the $k$-terms in  
Eq.~(\ref{twoloopconst})  
 (as well as the $R$-terms) 
are integrated over up to  
the resolution scale 
parameter $z_M$. 

\begin{figure}[htb]
\begin{center} 
\includegraphics[width=5cm]{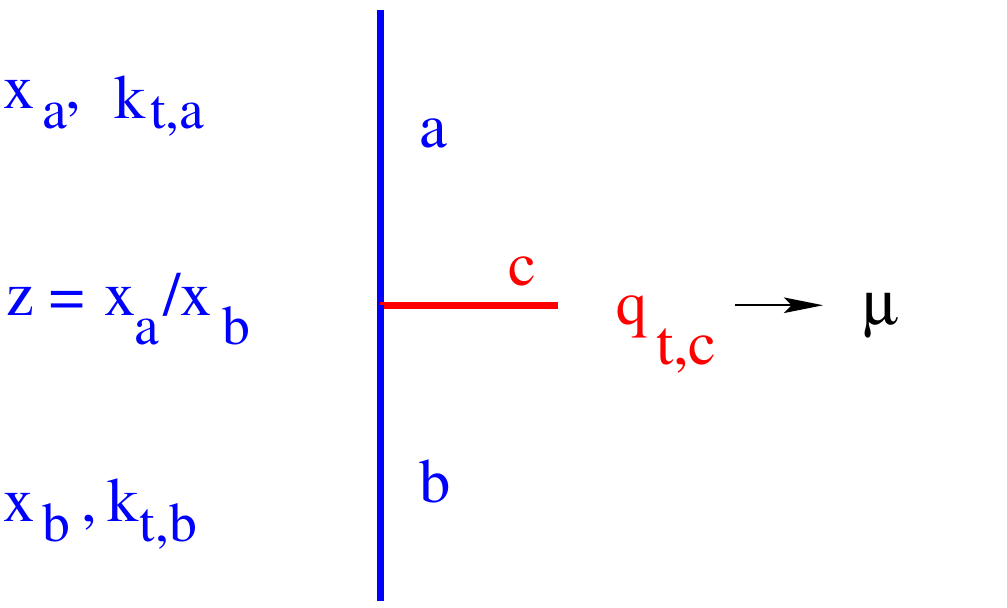}
  \caption{\it 
Branching 
process $ b \to a + c$.}
\label{fig:fig-kine}
\end{center}
\end{figure}

Next, we  solve Eq.~(\ref{sudint}) by 
numerical Monte Carlo method and 
use the parton branching 
kinematics (Fig.~\ref{fig:fig-kine}) 
to relate the transverse momentum 
recoils at each branching to the  evolution variable. With reference 
to  the notation of Fig.~\ref{fig:fig-kine} for the splitting $b \to a + 
c$, the plus lightcone momenta  are 
$p_a^+ = z p_b^+$,  
 $p_c^+ = (1-z) p_b^+$. 
 We  consider the cases of 
transverse-momentum ordering 
and angular ordering~\cite{skands05,gieseke03}.  
In the first case we have 
\begin{equation} 
\label{qtord} 
\mu = 
| {\bf q}_c | \, , 
\end{equation} 
where ${\bf q}_c $ is 
the (euclidean) transverse momentum 
vector of particle $c$.
In the second case we have 
\begin{equation} 
\label{angord} 
\mu = 
| {\bf q}_c | / (1 - z) \, . 
\end{equation} 
For  numerical   
solution,   
we develop a new program based 
on  the Monte Carlo  method 
which  was 
earlier employed 
by some of us 
for studies of the CCFM  
equations~\cite{Hautmann:2014uua,hj-updfs-1312}. 
The application    
to the case of the evolution equations 
(\ref{sudint})\footnote{First results from this numerical program  have been presented in~\cite{dis16proc}.}  presents different features with respect to the 
CCFM case. These  depend 
especially on the different 
flavor structure of the two 
 equations, and the different 
behavior of the kernels 
at small longitudinal momentum 
fractions. While CCFM equations 
are dominated by the gluon channel, 
Eq.~(\ref{sudint}) has  fully coupled flavor structure. The small-$x$ 
behavior of CCFM kernels is controlled by the non-Sudakov form factor.  
In the case    
of Eq.~(\ref{sudint}) it is essential to 
work with momentum-weighted distributions  to improve the  
convergence of the numerical  integration over the region of small $x$.     The  iterative solution 
of Eq.~(\ref{sudint}) 
schematically reads 
\begin{equation} 
\label{itera-a} 
  {\widetilde f}_a(x,\mu^2) = \sum_{i=0}^\infty {\widetilde f}^{(i)}_a(x,\mu^2)  \;\; , 
\end{equation}
where 
\begin{eqnarray} 
\label{itera-b} 
  {\widetilde f}^{(0)}_a(x,\mu^2) 
& = &  
S_a ( z_M, \mu^2 , \mu^2_0 ) \ 
 {\widetilde f}_a(x,\mu^2_0)    \; , 
\nonumber
\\  
{\widetilde f}^{(1)}_a(x,\mu^2) 
& = &  \sum_b 
\int^{\mu^2}_{\mu^2_0} 
{{d \mu^{\prime 2} } 
\over \mu^{\prime 2} } 
{
{S_a ( z_M, \mu^2 , \mu^2_0 )} 
 \over 
{S_a ( z_M, \mu^{\prime 2} , 
\mu^2_0 ) }
}
\int_x^{z_M} {dz} \;  {S_b ( z_M, \mu^{\prime 2} , 
\mu^2_0 ) }  \
\nonumber\\
& \times   & 
\left( 
K_{ab} (\as(\mu^{\prime 2})
) \ { 1 \over { 1 - z}} + 
R_{ab} (\as(\mu^{\prime 2})
,z) 
\right)
\ {\widetilde f}_b({x/z},
\mu_0^{ 2})  
   \; , \;\;  \dots \; . 
\end{eqnarray}
Using this branching 
Monte Carlo 
solution and the 
parton kinematic relations given 
above, 
 we are able to   compute    
the distribution ${\cal A}_a$ in the 
transverse momentum 
 $  {\bf k } = - \sum_c {\bf q}_c $, 
in addition to the inclusive 
 distribution, integrated 
over $  {\bf k }$,   
\begin{equation} 
\label{unintA}
\int  
x \ {\cal A}_a ( x , {\bf k } , \mu^2)  
\  { {d^2 {\bf k }} \over \pi} 
=  {\widetilde f}_a(x,\mu^2) \; . 
\end{equation}   

\begin{figure}[htb]
\begin{center} 
\includegraphics[width=7cm, trim=90 150 5 180, clip ]{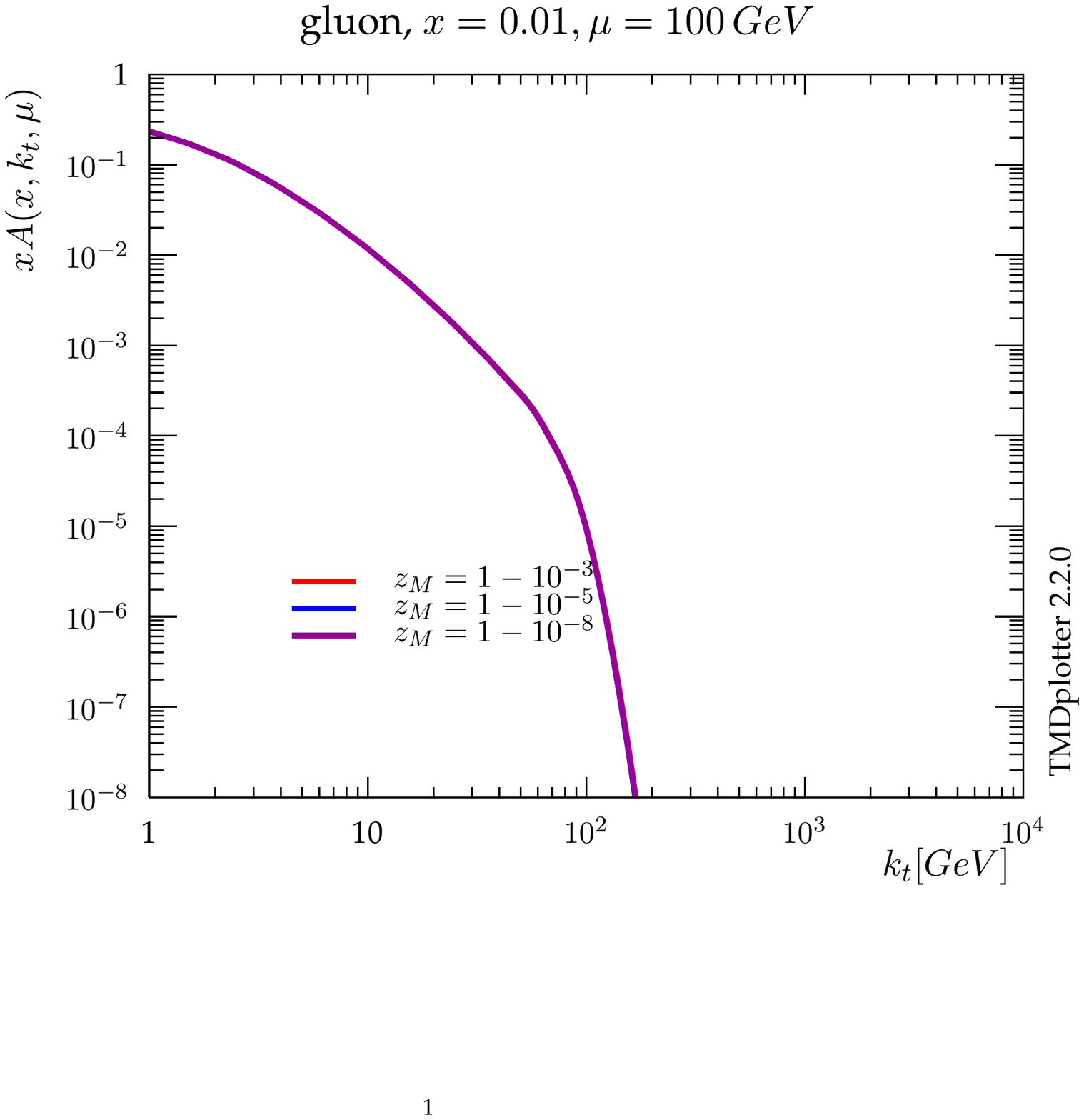} \hspace*{ 1.0cm}
\includegraphics[width=7cm, trim=90 150 5 180, clip ]{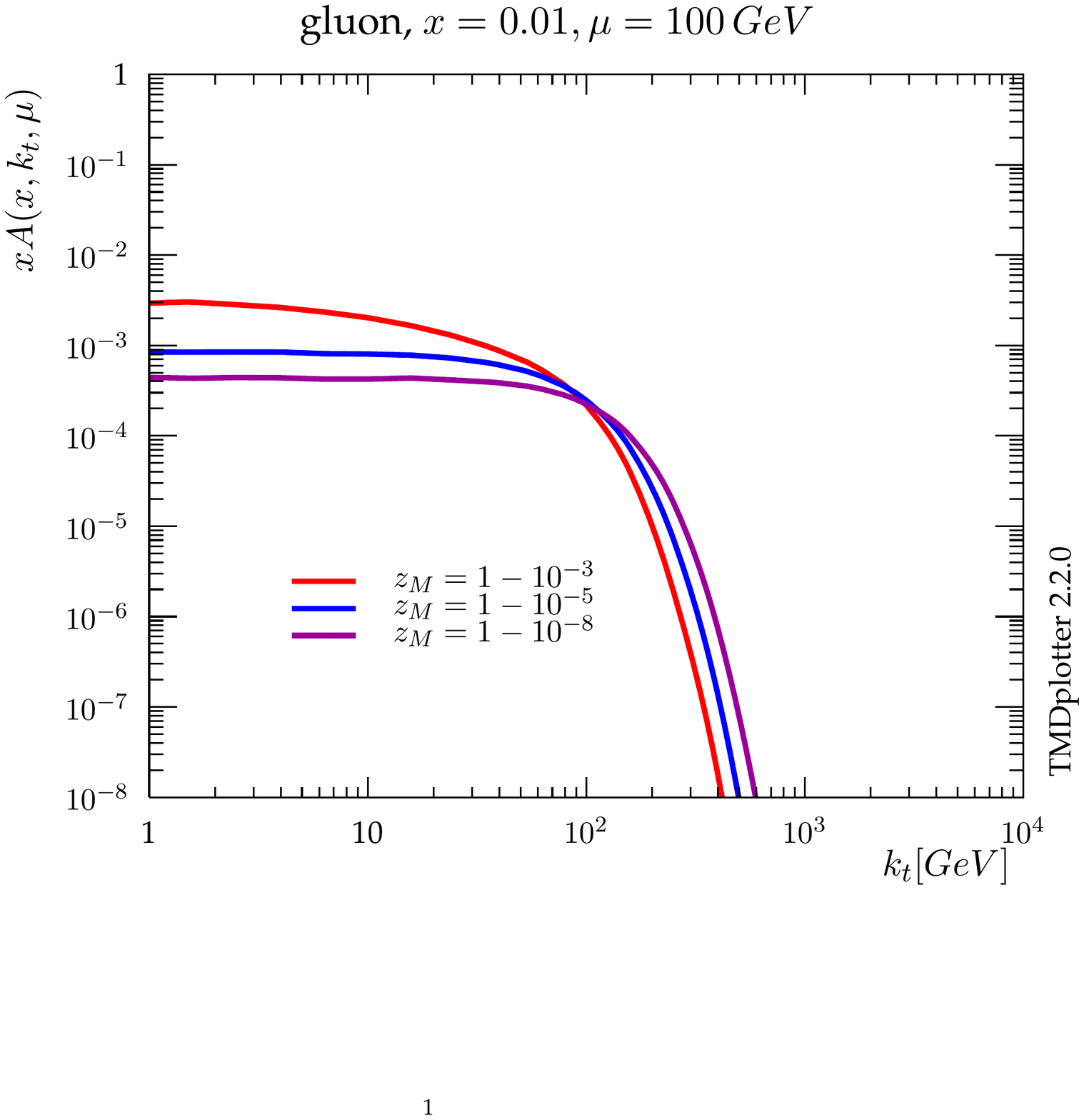}
\includegraphics[width=7cm, trim=90 150 5 180, clip ]{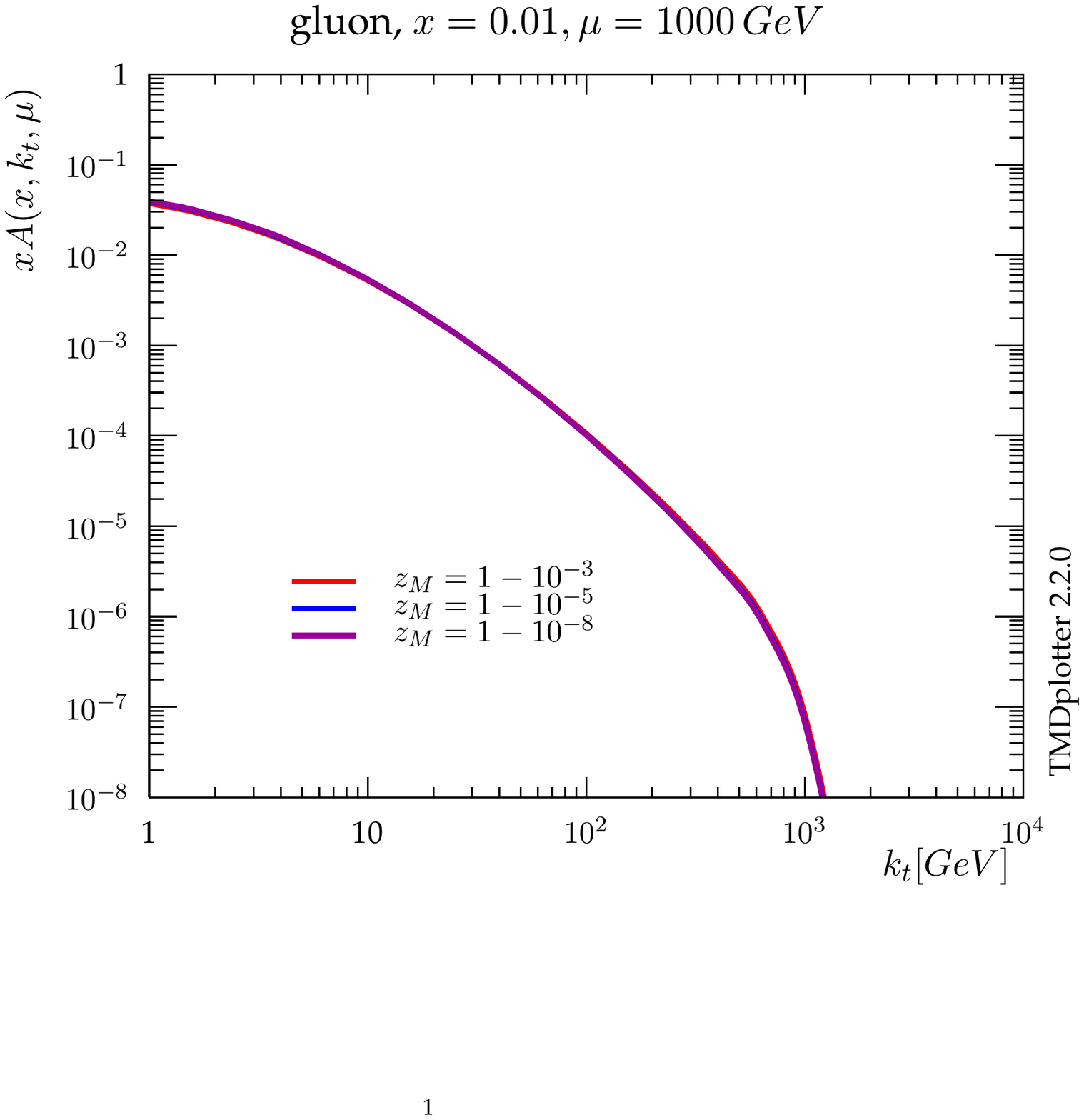} \hspace*{ 1.0cm}
\includegraphics[width=7cm, trim=90 150 5 180, clip ]{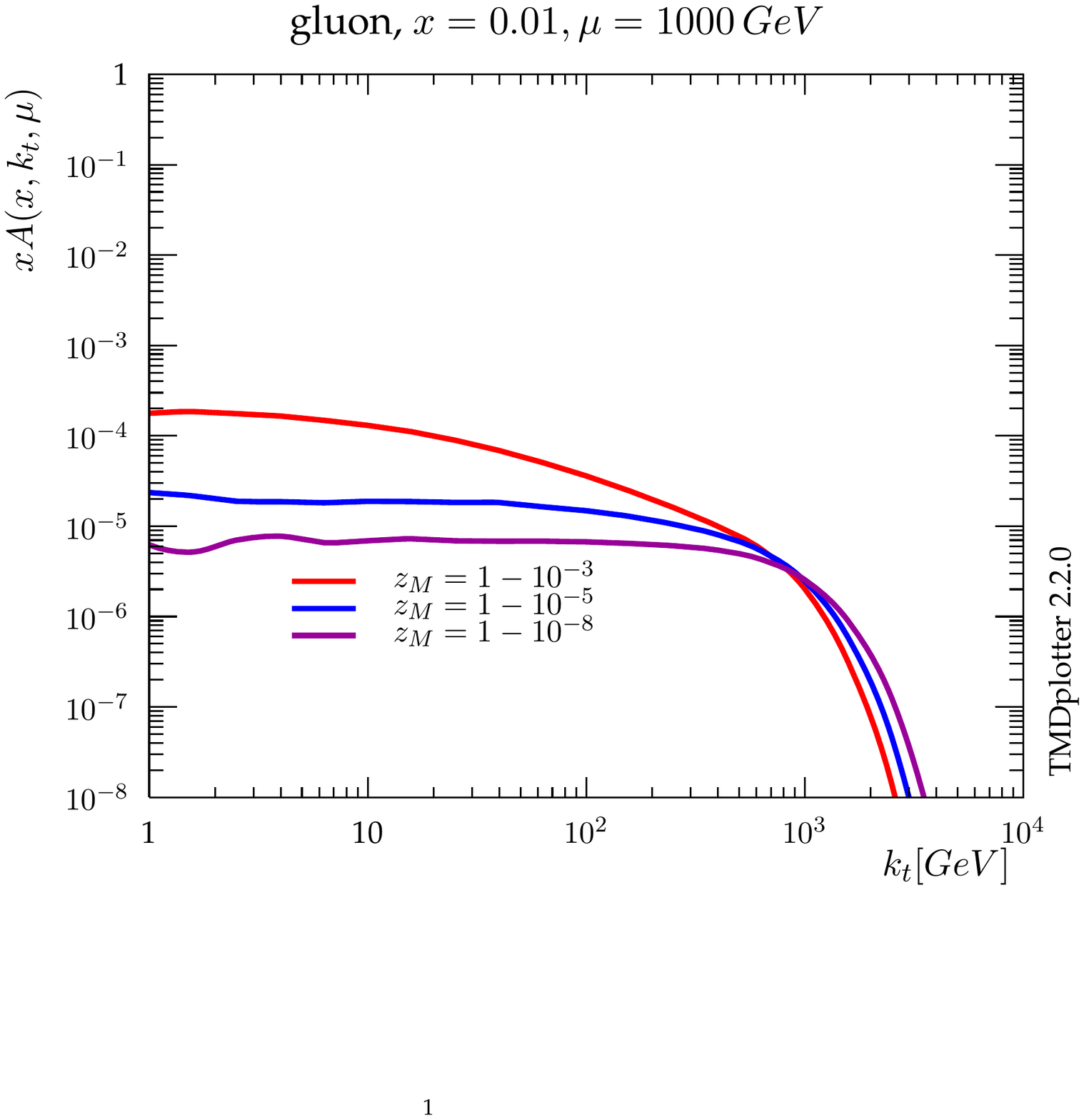}
  \caption{\it 
Transverse momentum 
gluon distribution 
 at $x=10^{-2}$ and  $\mu=100 
$~${\rm{GeV}}$ (upper row), $\mu = 1000$~${\rm{GeV}}$ (lower row) for different values of the resolution scale parameter $1 - z_{M}=10^{-3}, 10^{-5}, 10^{-8}$: 
(left) angular ordering; (right) transverse momentum 
ordering.}
\label{fig:fig-tmdzmax}
\end{center}
\end{figure}

The key  observation is that while 
in the case of the inclusive distribution 
the cancellation of real and virtual 
 non-resolvable emissions 
leads to results which become  independent 
of the resolution parameter $z_M$ for large enough $z_M$, regardless of the choice of the evolution variable,  the case of the transverse momentum 
distribution is infrared-sensitive and depends on the appropriate choice of the evolution variable (e.g.,    Eqs.~(\ref{qtord}),(\ref{angord})).  
In the framework of~\cite{fh07}, 
this infrared sensitivity is treated 
by using  the subtractive 
technique~\cite{Hautmann:2000cq} in the 
definition of transverse momentum 
dependent distributions and leads to a generalization of the plus distribution 
(\ref{plusdistrdef}). In the case of 
the branching solution of the evolution 
equations analyzed in this 
paper, we will see next that the 
angular ordering (\ref{angord}) 
takes into account the cancellation 
of non-resolvable emissions and gives stable, $z_M$-independent results.

\begin{figure}[htb]
\begin{center} 
\includegraphics[width=7cm, trim=90 150 5 180, clip ]{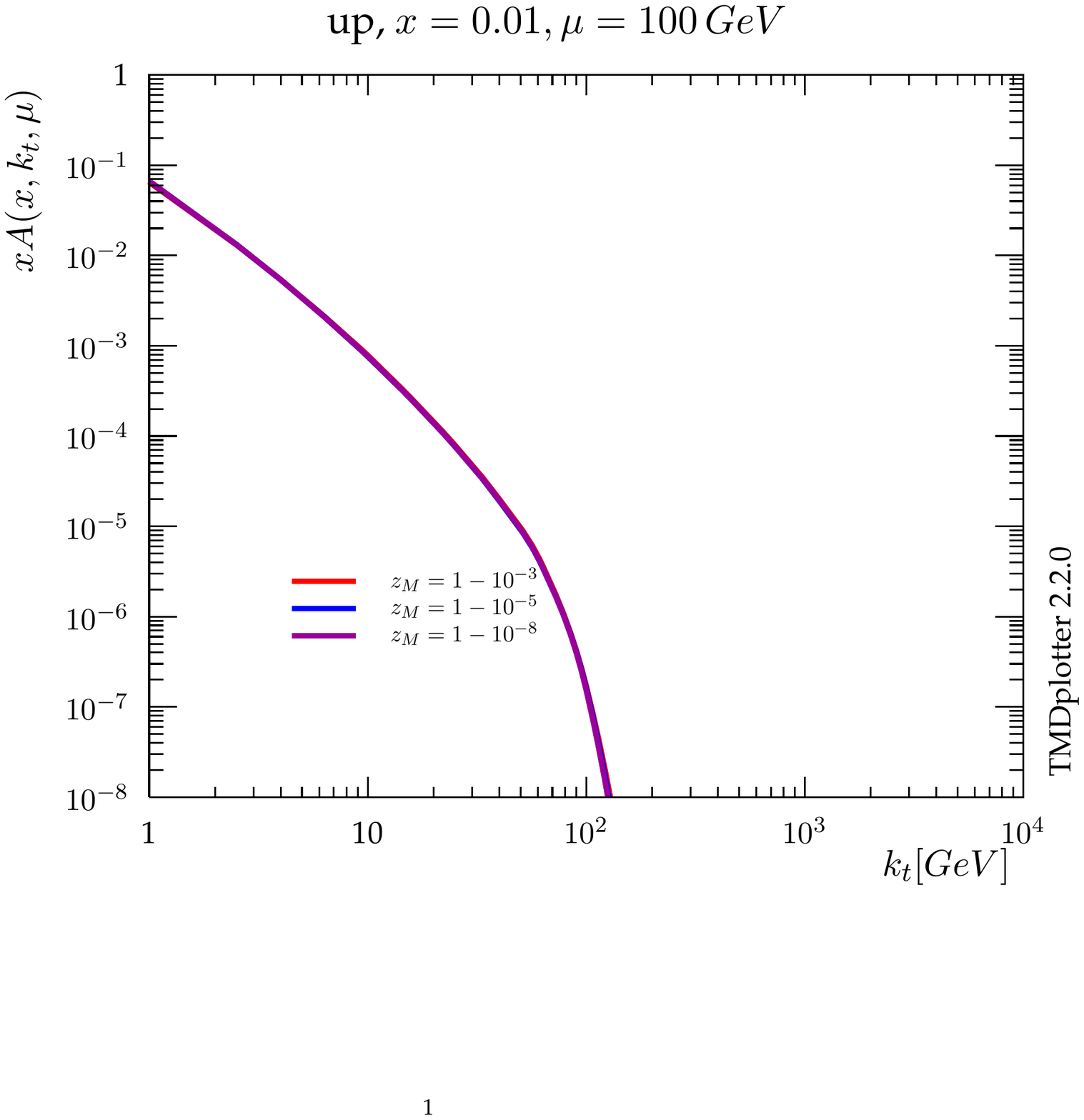} \hspace*{ 1.0cm}
\includegraphics[width=7cm, trim=90 150 5 180, clip ]{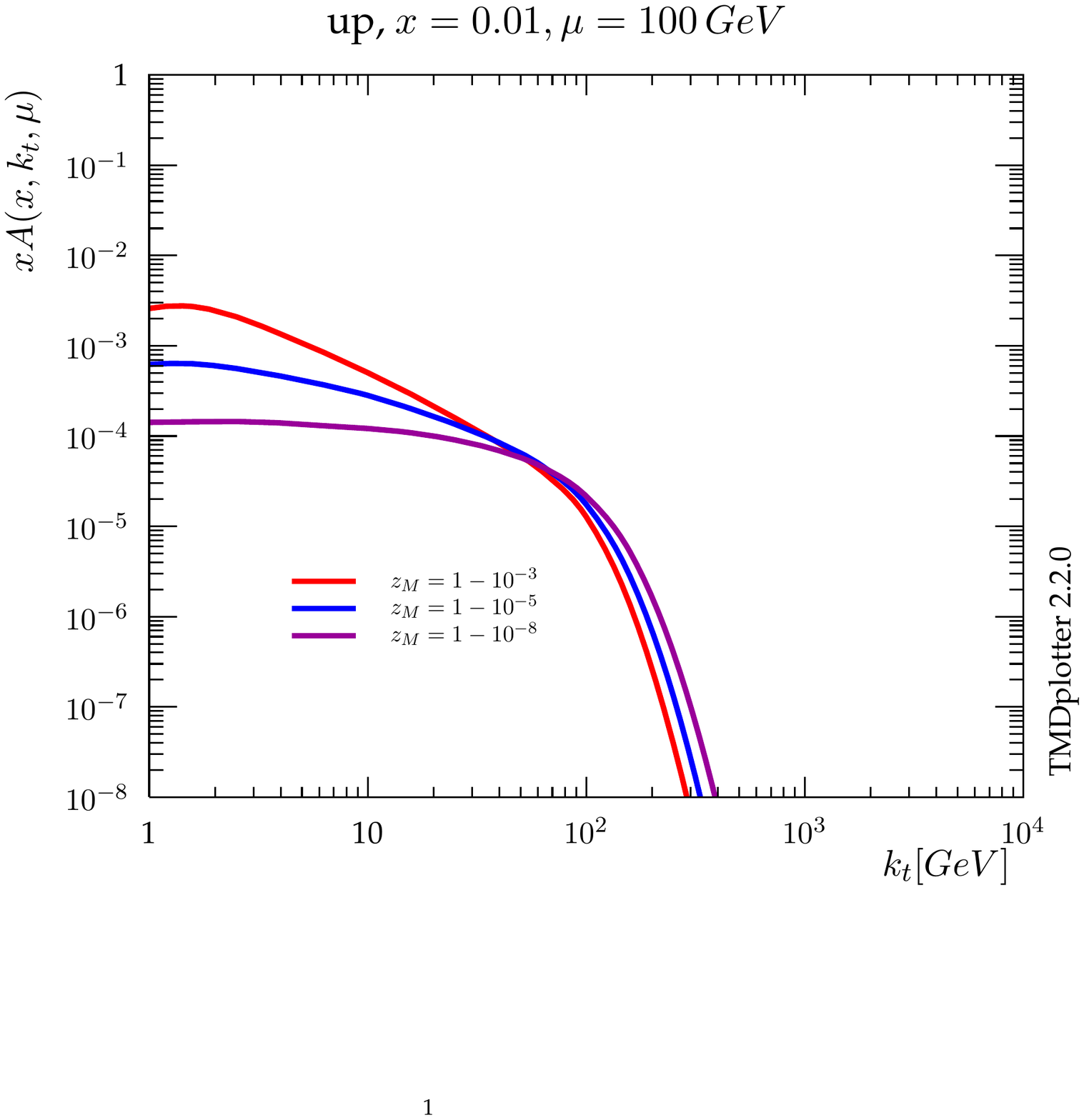}
\includegraphics[width=7cm, trim=90 150 5 180, clip ]{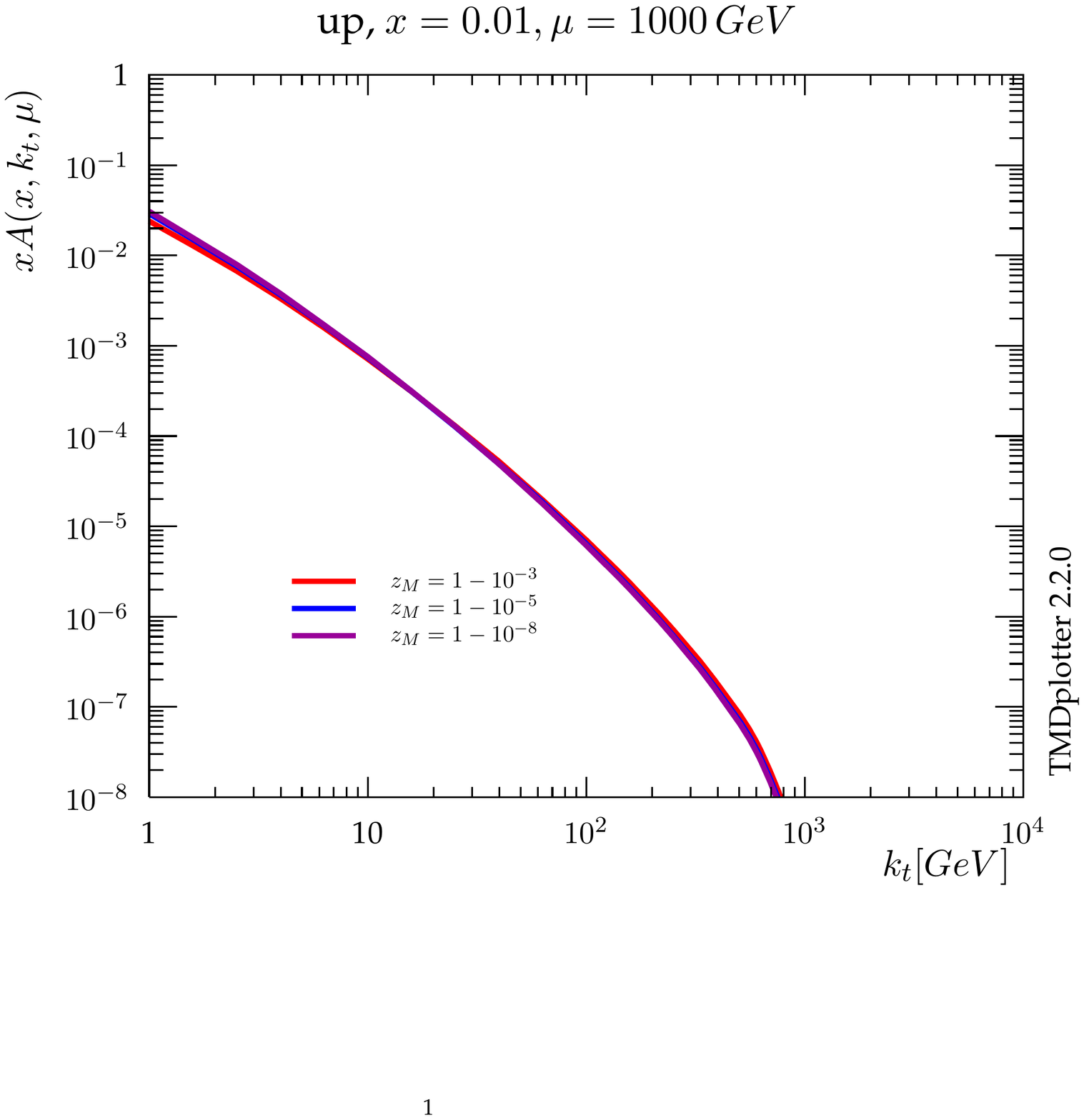} \hspace*{ 1.0cm}
\includegraphics[width=7cm, trim=90 150 5 180, clip ]{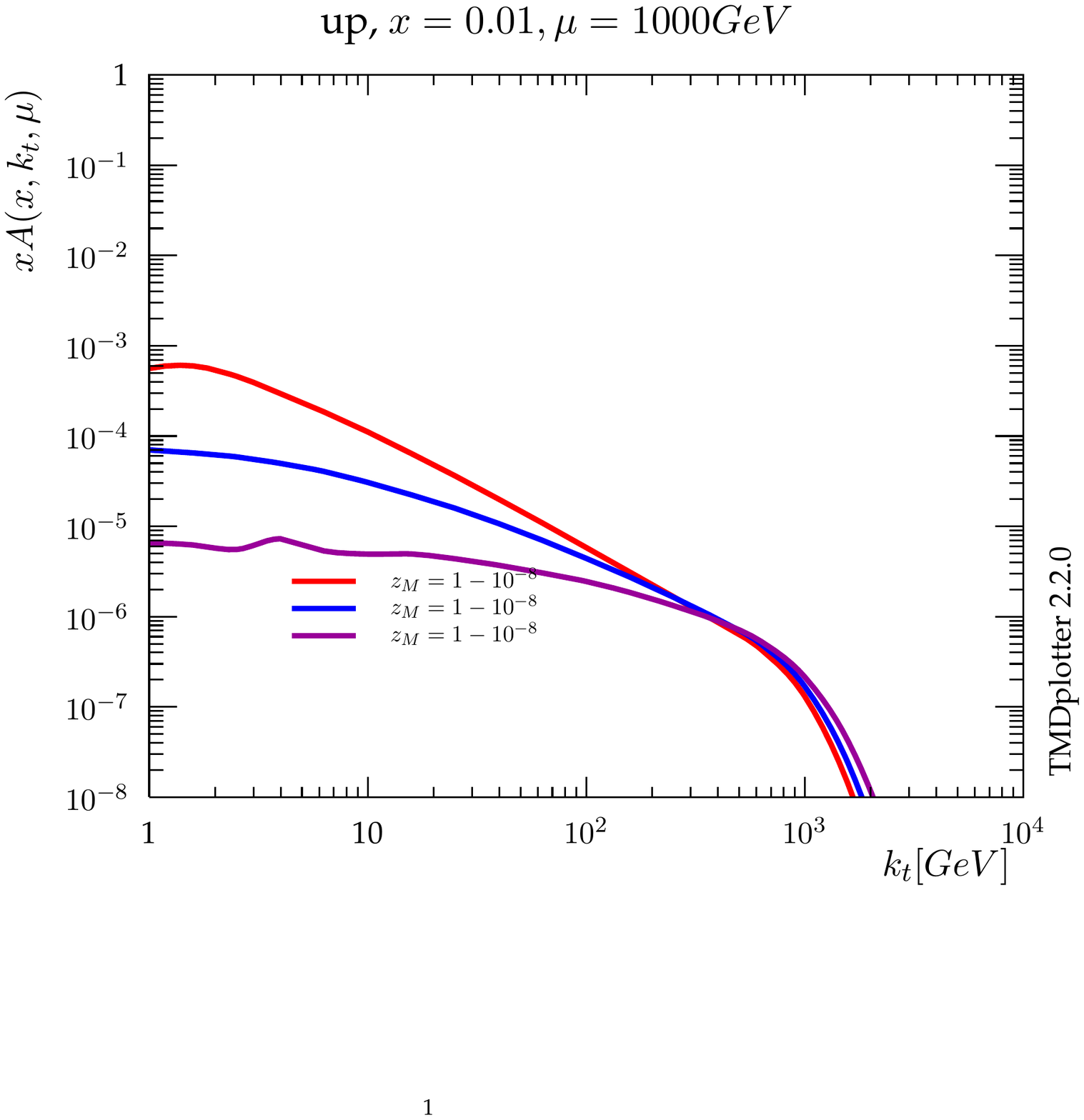}
  \caption{\it 
Transverse momentum 
up-quark distribution 
 at $x=10^{-2}$ and  $\mu=100 
$~${\rm{GeV}}$ (upper row), $\mu = 1000$~${\rm{GeV}}$ (lower row) for different values of the resolution scale parameter $1 - z_{M}=10^{-3}, 10^{-5}, 10^{-8}$: 
(left) angular ordering; (right) transverse momentum 
ordering.}
\label{fig:fig-tmdzmax-up}
\end{center}
\end{figure}

In Figs.~\ref{fig:fig-tmdzmax},\ref{fig:fig-tmdzmax-up} we 
apply  our  numerical  solution of 
Eq.~(\ref{sudint}) to study the  
transverse-momentum 
dependence of the gluon and up-quark distribution 
and their  behavior with the soft-gluon  resolution parameter  $z_M$.  
The parameter $z_M$ in general depends on the 
evolution scale $\mu$. For numerical illustrations in this
paper we limit ourselves to presenting results at fixed 
values of $z_M$.    Fig.~\ref{fig:fig-tmdzmax}  shows 
 the  gluon 
distribution versus $ k_t \equiv 
| {\bf k} | $ for different values of 
the resolution parameter, 
$1 - z_{M}=10^{-3}, 10^{-5}, 10^{-8}$. Fig.~\ref{fig:fig-tmdzmax-up} 
shows analogous curves for the up-quark distribution. The distributions are plotted for 
 a fixed value 
of longitudinal momentum fraction,  
$x=10^{-2}$, and two values 
of evolution scale, $ \mu = 100 
$~GeV (top panels) and 
$ \mu = 1000 
$~GeV (bottom panels).\footnote{The plots in 
Figs.~\ref{fig:fig-tmdzmax},\ref{fig:fig-tmdzmax-up} are produced using the plotting tool 
TMDplotter~\cite{tmdplott,tmdplott16}.} 
On the right  are the results for transverse-momentum ordering; on the left are the results for angular ordering. We see that the 
transverse-momentum ordering  
does not lead to results 
independent of $z_M$. In contrast, the angular ordering does. The different  
behavior  is associated with    
the emission of  gluons  at large  negative rapidities,  $ 
y \sim \ln ( p^+ / p^-)  \to  
- \infty $.  In the case of 
transverse-momentum ordering,  
for quantities which are not inclusive but depend on observed  transverse momenta,  
  an extra dependence is left over on $z_M$, corresponding to a cut-off on the rapidity of emitted gluons.
On the other hand, the angular 
ordering correctly 
takes into account the cancellation 
of non-resolvable emissions 
due  to soft-gluon 
coherence~\cite{Catani:1990rr}, and 
no dependence 
is left on the resolution scale. 

\begin{figure}[htb]
\begin{center} 
\includegraphics[width=7.8cm]{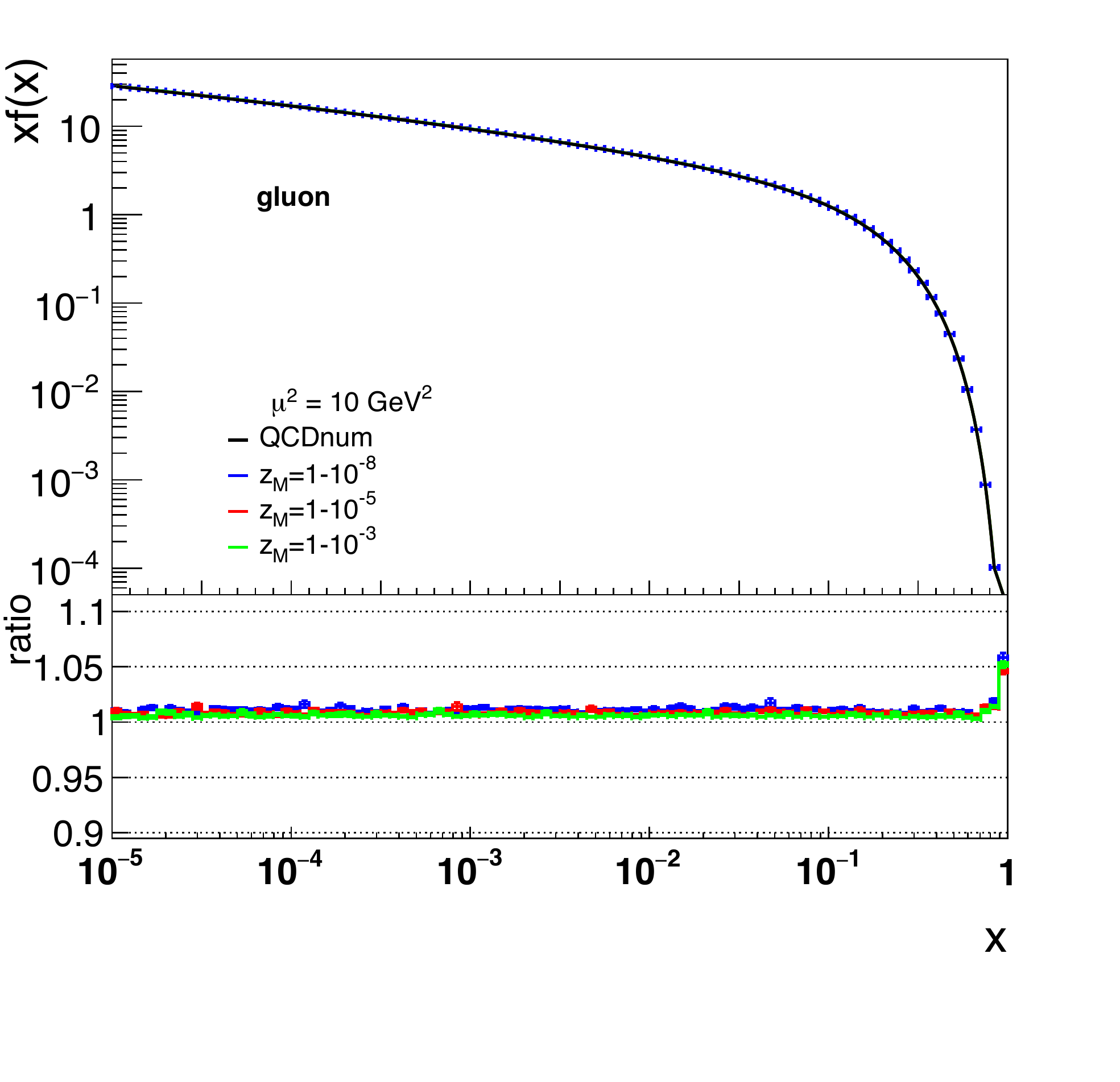}\includegraphics[width=7.8cm]{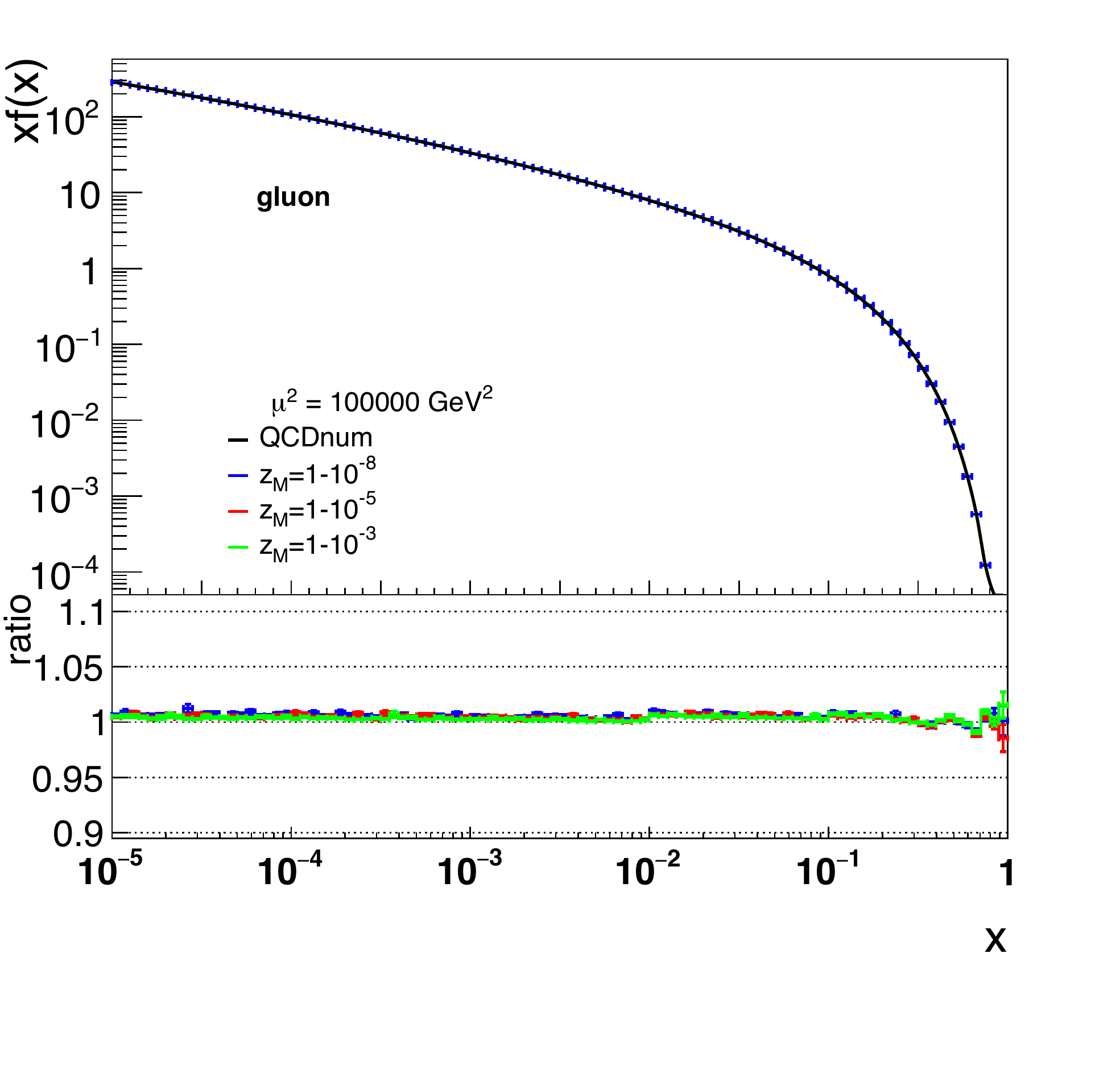}
\includegraphics[width=7.8cm]{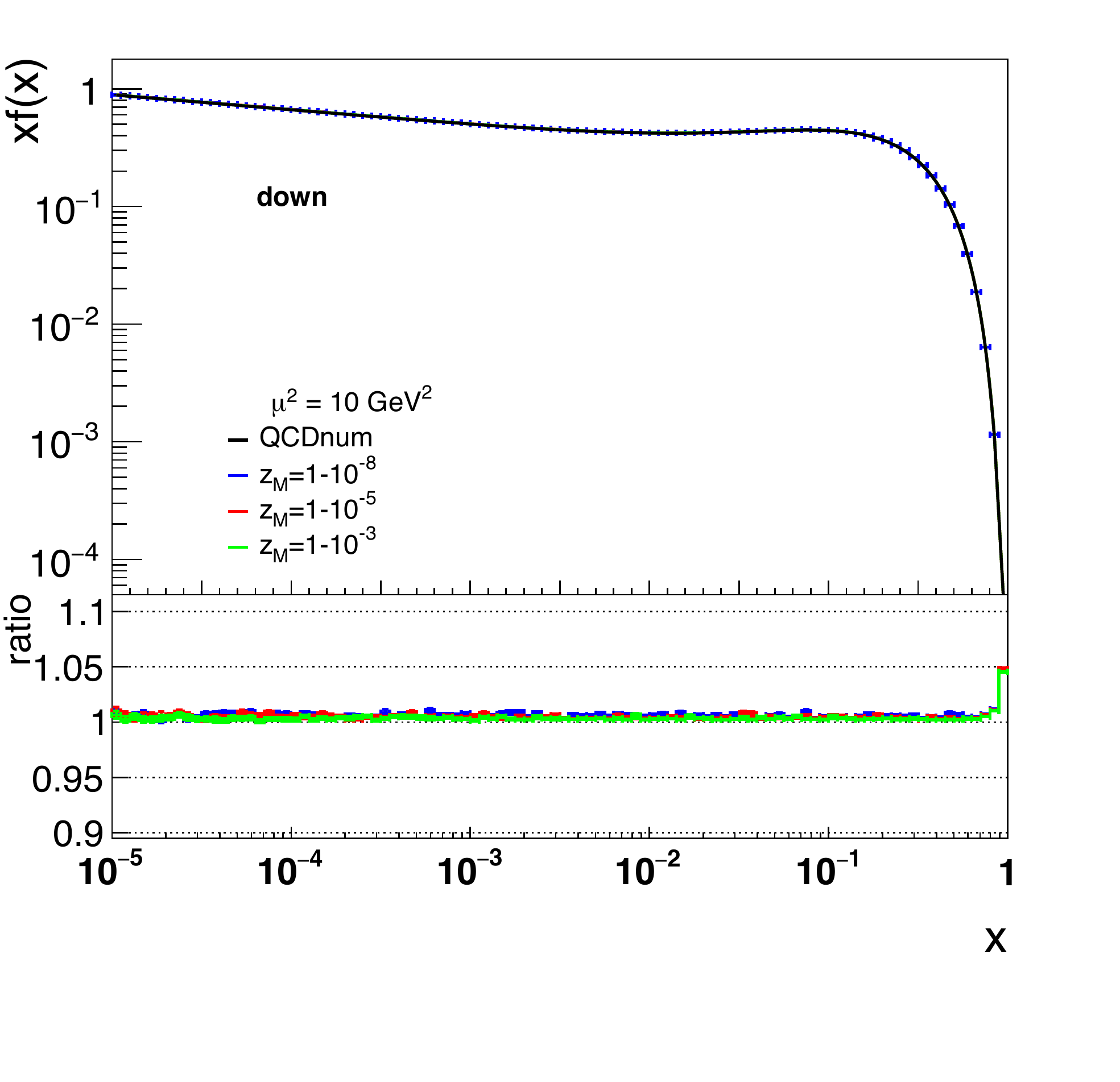}\includegraphics[width=7.8cm]{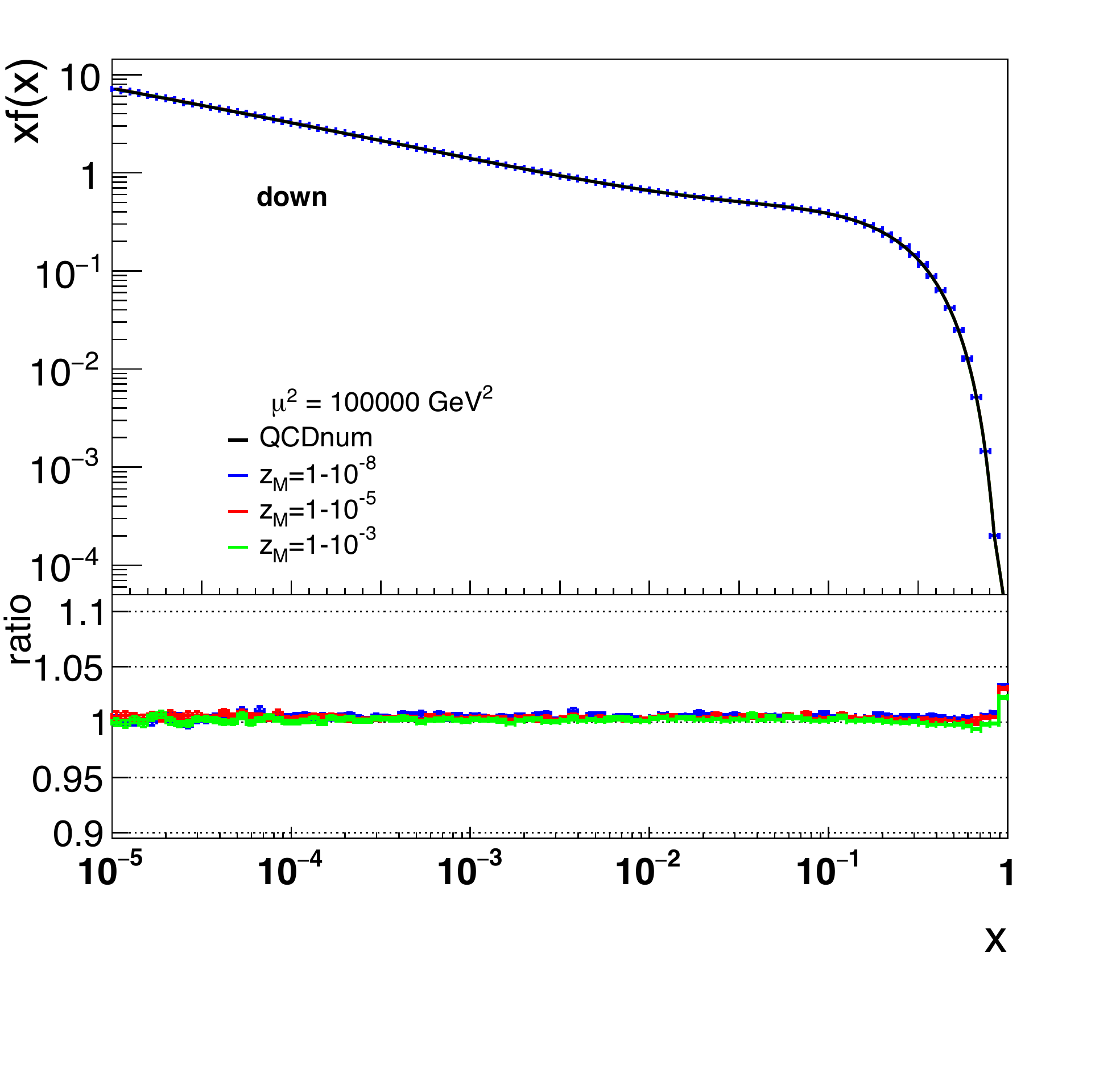}
  \caption{\it 
Integrated 
gluon and down-quark distributions  
 at   $\mu^2=10   
$~${\rm{GeV}}^2$ (left column) and $\mu^2 = 10^{5}$~${\rm{GeV}}^2$ (right column) obtained 
from the parton-branching  solution for different values of  
$ z_{M}$, 
compared with the result from 
{\sc Qcdnum}. The ratio plots show the ratio of the results obtained 
with the parton-branching  method to  the result from {\sc Qcdnum}.}
\label{fig:fig-qdcnum}
\end{center}
\end{figure}

We find  that the effect from the ordering variable 
and soft-gluon resolution  illustrated in 
Figs.~\ref{fig:fig-tmdzmax},\ref{fig:fig-tmdzmax-up}  
influences the transverse momentum 
distribution for any  
kinematic region, namely, both at 
small $k_t$ and large $k_t$, and  
for any value of $x$  (including 
the low-$x$ region).

The result in 
Figs.~\ref{fig:fig-tmdzmax},\ref{fig:fig-tmdzmax-up}   
is obtained using an arbitrarily 
chosen   
form  for the 
distributions  at the initial scale 
of evolution $\mu_0$ 
in Eq.~(\ref{itera-b}). 
This is sufficient to illustrate 
the main point about the 
$z_M$ dependence. 
In a complete 
treatment, the initial distributions 
are to be determined from fits to 
experimental data. We 
plan to   report on this in a 
future publication. 

Also, we have obtained the 
numerical curves  in 
Figs.~\ref{fig:fig-tmdzmax},\ref{fig:fig-tmdzmax-up}   
by restricting ourselves to  
leading order in the strong 
coupling. The method described 
in this paper however is general 
and can be extended to higher 
orders. Explicit results at the  
next-to-leading order  will be presented  in a separate paper.

Fig.~\ref{fig:fig-qdcnum} shows the 
result of applying our 
parton-branching  solution of 
Eq.~(\ref{sudint}) to compute the evolution of 
gluon and quark distributions   
as functions of $x$, for different 
values of the resolution scale 
parameter $z_M$. We use this to validate our 
method in the case of 
ordinary parton distributions, integrated over $ k_t$. As a 
consistency check we 
verify that, regardless of the ordering variable in 
Fig.~\ref{fig:fig-tmdzmax}, the 
result for the inclusive parton distributions converges as a function 
of $z_M$ for large enough $z_M$. 
Further, we  compare the answer 
from our 
parton-branching solution 
of the evolution equations 
to semi-analytic  results  obtained 
via the evolution package 
{\sc Qcdnum}~\cite{Botje:2010ay,qcdnum-pre1,qcdnum-pre2}.\footnote{Similar 
comparisons were made 
in~\cite{Jadach:2003bu,GolecBiernat:2006xw} 
and 
in~\cite{tanaka03}.}   We find 
agreement to a  level better than 
1 $\%$.  


 In conclusion, we have shown that 
the evolution of parton distribution 
functions can be calculated, including 
the transverse momentum dependence, 
from a parton branching approach,  provided infrared contributions  are treated by a method which takes into account 
consistently soft gluon emissions near the endpoint $ z \to 1$ not just at inclusive level but at exclusive level. 
We have analyzed in detail the dependence on the soft-gluon resolution scale parameter $z_M$. 

Having defined this properly  opens 
the way to collider applications of 
TMD parton distributions. Contrary to most studies so far, which are limited to specific kinematic regions, the approach of  this paper is expected to be valid more generally and in particular includes the full flavor structure. 

\vskip 0.9 cm 

\noindent {\bf Acknowledgments}. We are grateful to 
S.~Jadach for many fruitful discussions on 
the topics of this paper. 
FH  acknowledges the support and hospitality of DESY, the 
University of Hamburg and the DFG Collaborative Research Centre SFB 676 
``Particles, Strings and the Early Universe".


\end{document}